\documentclass[prl,reprint,superscriptaddress]{revtex4-1}

\usepackage{mathtools,bm,amsfonts,color, graphicx, subfigure,tikz}
\usetikzlibrary{arrows,automata}

\begin{document}
\title{Graph-Facilitated Resonant Mode Counting in Stochastic Interaction Networks}
\author{Michael F Adamer}
\affiliation{Wolfson Centre for Mathematical Biology, Mathematical Institute, University of Oxford}
\author{Thomas E Woolley}
\affiliation{Wolfson Centre for Mathematical Biology, Mathematical Institute, University of Oxford}
\author{Heather A Harrington}
\affiliation{Wolfson Centre for Mathematical Biology, Mathematical Institute, University of Oxford}

\date{\today}
\begin{abstract}
Oscillations in a stochastic dynamical system, whose deterministic counterpart has a stable steady state, are a widely reported phenomenon. Traditional methods of finding parameter regimes for stochastically-driven resonances are, however, cumbersome for any but the smallest networks. In this letter we show by example of the Brusselator how to use real root counting algorithms and graph theoretic tools to efficiently determine the number of resonant modes and parameter ranges for stochastic oscillations. We argue that stochastic resonance is a network property by showing that resonant modes only depend on the squared Jacobian matrix $J^2$, unlike deterministic oscillations which are determined by $J$.  By using graph theoretic tools, analysis of stochastic behaviour for larger networks is simplified and  chemical reaction networks with multiple resonant modes can be identified easily.
\end{abstract}
\maketitle

\textit{Introduction.---}Interaction networks are ubiquitous in biological physics and mathematics \cite{Murray2002,Murray2008,Kampen2007}, from predator-prey models \cite{McKane2005,Wang1997,Yoshida2003,Lugo2008,Lotka1925} to the vast field of chemical reaction networks \cite{Feinberg1987,Feinberg1988,Mincheva2007,Michaelis1913}. Previous research highlighted how resonant amplification of noise in stochastic interaction networks can lead to behaviour not anticipated from deterministic ordinary differential equation (ODE) models, in particular the emergence of cyclic behaviour in stochastic models where the deterministic counterpart does not show a Hopf bifurcation \cite{McKane2005}. 

The main tools for investigating stochastic cycles are based on the calculation of the exact power spectra for the constituents of the network from a Langevin equation \cite{Woolley2011,Kampen2007,Simpson2004}, which demands knowledge of noise covariances. The determination of noise covariance requires extensive coarse graining, starting from a master equation formulation of the interaction system, and via weak noise expansions the deterministic equations, and a Fokker-Planck equation can be calculated. Eventually coarse graining allows the use of the simpler chemical Langevin equation \cite{Kampen2007}. We seek to streamline the coarse graining process by showing how the desired information, namely the number of resonant frequencies of a network, can be extracted from the deterministic equations only. We also find the parameter ranges associated with a number of resonant modes using graph theoretical approaches developed for chemical reaction networks.

There is a large body of algebraic and graph theoretic techniques for studying deterministic mathematical models. Usually these interaction networks have a large number of parameters, typically one rate constant per interaction and the model parameters are responsible for the dynamics of the system \cite{Guldberg1864,Feinberg1987}. Past research focussed successfully on exploiting the network structure of an interaction system for determining its dynamical behaviour, as network structure is a feature of a model and unaffected by the choice of rate constants \cite{Feinberg1987,Feinberg1988,Mincheva2007}. In \cite{Feinberg1987} it was shown how network structure can be used to determine whether a given chemical reaction network has stable steady states, a useful tool to rule out multistationarity in a network. More recently graph theoretical methods have been employed to show how network features such as feedback cycles can lead to oscillations and multistationarity in chemical reaction networks \cite{Mincheva2007}. Graph theoretical methods provide the additional advantage over the approach in \cite{Feinberg1987} that they allow one to explore the bifurcation structure of the network. Despite the apparent advantage of using graph theoretical methods for the investigation of dynamical capabilities of interaction networks the graph based investigation of stochastic models is still in its infancy \cite{Meng2015}. 

In this letter we provide an alternative route for calculating the resonant frequencies (and its parameter range) of stochastically-driven oscillating systems. Instead of solving the roots of a rational function of the power spectrum from the weak noise approximation, we investigate the maxima of this function. To do this, we adapt algebraic techniques (e.g. Sturm chains) and a graph theoretic formulation for finding the coefficients of the characteristic polynomial and thereby offering a methodology for studying stochastically-driven oscillations without requiring excessive expansions.

\textit{Weak noise and power spectrum.---}The Brusselator is a model for an autocatalytic reaction such as the Belousov-Zhabotinsky reaction \cite{nicolis1977} and follows the reaction scheme
\begin{align*}
A&\rightarrow X,\\
2X+Y&\rightarrow 3X,\\
B+X&\rightarrow Y+D,\\
X&\rightarrow \emptyset.
\end{align*}
In the Brusselator model the chemical species $A$ and $B$ are assumed to be constant and hence represent the model parameters.
Using the stochastic version of the law of mass action \cite{Anderson2010} we can determine the reaction rates
\begin{align}
T(X+1,Y|X,Y) &= A\Omega,\nonumber\\
T(X-1,Y|X,Y) &= X,\nonumber\\
T(X+1,Y-1|X,Y) &= \frac{X(X-1)Y}{\Omega^2},\nonumber\\
T(X-1,Y+1|X,Y) &= BX,
\end{align}
where $\Omega$ represents the total volume of the system.
To formulate the chemical master equation we define the operators $E_{X}^\pm$ and $E_{Y}^\pm$ on functions of $X$, $Y$ and time $t$ as
\begin{align}
E_X^\pm f(X,Y,t) &= f(X\pm 1,Y,t),\nonumber\\
E_Y^\pm f(X,Y,t) &= f(X,Y\pm 1,t).
\label{LadderOps}
\end{align}
The definition in \eqref{LadderOps} allows us to write the master equation for the Brusselator in the compact form
\begin{align}
\frac{d P(X,Y,t)}{dt} &= (E_X^+ - 1)T(X-1,Y|X,Y)P(X,Y,t)\nonumber\\
&+(E_X^- - 1)T(X+1,Y|X,Y)P(X,Y,t)\nonumber\\
&+ (E_X^+E_Y^- - 1)T(X-1,Y+1|X,Y)P(X,Y,t)\nonumber\\
&+ (E_X^- E_Y^+ - 1)T(X+1,Y-1|X,Y)P(X,Y,t).
\label{MasterEq}
\end{align}
For large $\Omega$ a Van Kampen expansion \cite{Kampen2007} of equation \eqref{MasterEq} is of the form
\begin{equation}
\frac{X}{\Omega} = u_1(t) + \frac{x_1}{\sqrt{\Omega}},
\end{equation}
with $x_1$ as a new stochastic variable. A similar expansion for $Y$ yields the deterministic equations for the Brusselator
\begin{align}
\dot{u}_1 &= A + u_1^2u_2 - (B+1)u_1,\nonumber\\
\dot{u}_2 &= Bu_1 - u_1^2u_2.
\label{Brusselator}
\end{align}
It is well known that the Brusselator exhibits a supercritical Hopf bifurcation when $A^2+1=B$ \cite{nicolis1977} and that the system has a stable steady state at $(A, B/A)$ when $A^2+1 < B$, which is the focus of our analysis.
The deterministic equations represent the leading order of the expansion in the limit where $\Omega$ is large and at the next order we obtain a Fokker-Planck equation \cite{Kampen2007}. At steady state it is, however, simpler to use the equivalent representation of a chemical Langevin equation \cite{Kampen2007,McKane2005}
\begin{align}
\bm{\dot{x}} = J\bm{x} + \bm{\lambda},
\label{Langevin}
\end{align}
where bold quantities represent vectors, $J$ is the Jacobian of \eqref{Brusselator} evaluated at the fixed point,
\begin{equation}
J = \begin{pmatrix}
B-1 & A^2\\
-B & -A^2
\end{pmatrix},
\end{equation}
and $\bm{\lambda}$ is a vector of Gaussian Markov processes. Equation \eqref{Langevin} determines the stochastic behaviour of the Brusselator at large, but finite $\Omega$.


A useful tool to find oscillations in stochastic trajectories is the power spectrum $P_k(\omega^2) = \langle |\hat{x}_k|^2\rangle$ where $\hat{x}_k$ is the Fourier transform of the $k^{\text{th}}$ element of \eqref{Langevin} and $\langle\cdot\rangle$ denotes the average over a number of realisations \cite{Woolley2011}. The general form of the power spectrum of the $k^{\text{th}}$ species of any interaction network whose stochastic behaviour can be described by equation \eqref{Langevin} is
\begin{align}
P_k(\omega^2) &= \frac{Q_k(\omega^2)}{R(\omega^2)},\label{PSpec}\\
\intertext{with}
R(\omega^2) &= \text{det}(J^2 +\omega^2 I),\label{R}\\
Q_k(\omega^2) &= \langle [\text{adj}(J+i\omega)\hat{\bm{\lambda}}]_k\;[\text{adj}(J-i\omega)\hat{\bm{\lambda}}]_k\rangle,
\label{Q}
\end{align}
where $I$ is the identity matrix, $\text{adj}(\cdot)$ is the adjugate matrix, $\text{det}(\cdot)$ is the determinant and $\langle\cdot\rangle$ denotes the average. $R(\omega^2)$ and $Q_k(\omega^2)$ are polynomials of degree $n$ and $n-1$ with $n$ being the number of species in the network, in the case of the Brusselator $n=2$. Note that $R(\omega^2)$ reduces to the characteristic polynomial of $J^2$ if we let $\omega^2 = -\lambda$. {Previous approaches proceeded by analysing all $n$ rational functions \eqref{PSpec} to determine the exact shape of the power spectra, and hence prove the existence of maxima. We will show how to determine the number of peaks and their parameter ranges by considering a single polynomial equation.} Stochastic oscillations manifest themselves as peaks in the power spectra. From equation \eqref{PSpec} it becomes apparent that peaks may either arise either from maxima of $Q_k(\omega^2)$ or minima of $R(\omega^2)$ or both. In analogy with the damped harmonic oscillator we define $\omega_R$ as a resonance frequency or resonant mode such that $R(\omega_R^2)$ is a minimum. Our definition implies further that the resonance frequencies are properties of the underlying network structure, represented by $J^2$, rather than the individual network constituents. Surprisingly, the number of resonant modes is independent of the noise covariances $\langle \lambda_i\lambda_j\rangle$, even though resonance in interaction networks is a stochastic effect, giving further indication that resonance is a network property.

\noindent \textit{Sturm chains for counting the maxima of power spectrum.---}We now turn to determine the number of resonant modes in a given network and show how parameter ranges for stochastic oscillations can be calculated in the Brusselator.
At resonance the polynomial $R(\omega^2)$ has a minimum which translates into the condition
\begin{equation}
\frac{\text{d} R(\omega^2)}{\text{d}(\omega^2)} = R'(\omega^2) = 0
\label{Cond1}
\end{equation}
and, since the angular frequency $\omega$ is a real number, we are interested in finding all distinct, real, positive solutions to equation \eqref{Cond1}. A method to determine an upper bound of such solutions is given by `Descartes' rule of signs' \cite{struik2014} which states that the maximum number of real, positive roots of a polynomial is given by the number of sign changes of consecutive non-zero coefficients, if the terms of the polynomial are ordered with descending variable exponent. Descartes' rule, however, only gives an upper bound and counts multiple roots as distinct roots.

An exact root counting algorithm is given through the computation of Sturm sequences and the use of Sturm's theorem \cite{Sturm1829}. For a univariate polynomial $p(x)$ Sturm's theorem gives the number of distinct real roots in an interval $(a,b]$ with $a < b$. To apply Sturm's theorem we compute a Sturm chain for $p(x)$
\begin{align}
p_0 &= p(x),\nonumber\\
p_1 &=  \tfrac{dp(x)}{dx} = p'(x),\nonumber\\
p_2 &= -\text{rem}(p_0,p_1),\nonumber\\
&\vdots\nonumber\\
p_i &= -\text{rem}(p_{i-1},p_{i-1}),\nonumber\\
&\vdots\nonumber\\
0 &= -\text{rem}(p_{m-1},p_m),
\end{align}
where $\text{rem}(\cdot,\cdot)$ is the remainder of the polynomial long division. Sturm's theorem proceeds by considering the signs of the Sturm chain $p_0,p_1,\cdots,p_m$ evaluated at the points $a$ and $b$. Similarly to Descartes' rule the number of sign changes of $p_0(a),p_1(a),\cdots,p_m(a)$ and $p_0(b),p_1(b),\cdots,p_m(b)$ is counted which we denote as $\sigma(a)$ and $\sigma(b)$. The number of distinct real roots is simply  $\sigma(a) - \sigma(b)$. Letting $a=0$ and $b =\infty$ gives the number of all positive, distinct, real roots. For small networks, especially the case $n=2$, the number of real roots follows trivially from the quadratic formula and $\text{det}(A+xI) = x^2 + \text{Tr}(A)x+\text{det}(A)$, where $Tr(A)$ is the trace. When turning to larger networks, however, Sturm chains become an invaluable tool.

\begin{figure}
\centering
\includegraphics[scale=0.65]{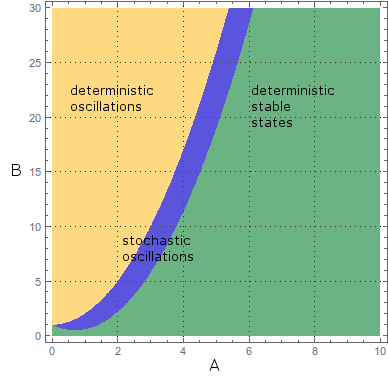}
\caption{The phase diagram for the stochastic Brusselator. In the stochastic weak noise regime there exists a band (blue, colour online) between the stable oscillations and the stable steady state where stochastic oscillations can be seen.}
\label{StochOsc}
\end{figure}

Returning to the Brusselator, $R(\omega^2)$ is given by
\begin{equation}
R(\omega^2) = \omega^4 + [(B-1)^2 + A^2(A^2 - 2B)]\omega^2 + A^4
\end{equation}
for which we can build the Sturm chain
\begin{align}
p_0 &= 2\omega^2 + (B-1)^2 + A^2(A^2 - 2B),\nonumber\\
p_1 &= 2,\nonumber\\
p_2 &= 0,
\label{BrussSturm}
\end{align}
hence,
\begin{align}
\sigma(0) &=
\begin{cases}
0\;\;\;\; \text{if } (B-1)^2 + A^2(A^2 - 2B) > 0\\
1\;\;\;\; \text{if } (B-1)^2 + A^2(A^2 - 2B) < 0
\label{BrusslCond1}
\end{cases},
\intertext{and}
\sigma(\infty) = 0.
\end{align}
Hence, necessary and sufficient conditions for the Brusselator to show stochastic oscillations are
\begin{subequations}
\begin{align}
&B < 1+A^2 \text{ (steady state condition),}\\
&(B-1)^2 + A^2(A^2 - 2B) < 0 \text{ (peak condition).}
\end{align}
\end{subequations}
This system of inequalities can be solved in a computational mathematics software such as Mathematica to give the  region of stochastic oscillations shown in Figure \ref{StochOsc}. 

From equations \eqref{BrussSturm} and \eqref{BrusslCond1} it becomes apparent that often we only need to evaluate specific coefficients of $R(\omega^2)$ rather than find the polynomial itself. Often, unless exact parameter ranges are needed, even fewer polynomial coefficients need to be considered due to some coefficients' inability to change sign, a feature easily identified from network motives in the graph of $J^2$. In the remainder of this letter we will outline a graph-based method to facilitate the finding of coefficients of $R(\omega^2)$ based on \cite{Mincheva2007}. 

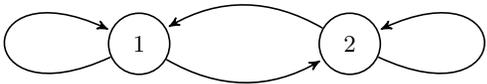
\begin{figure}
\centering
\begin{tikzpicture}[->,>=stealth',shorten >=1pt,auto,node distance=2.8cm, semithick]
  \tikzstyle{every state}=[fill=none,draw=black,text=black]
  \node[state] (A) {$1$};
  \node[state] (B)[right of = A] {$2$};
  
  \path (A) edge[bend right] node[below] {} (B)
  		(B) edge[bend right] node[above] {} (A);
  \path (A) edge[loop left, looseness=15, in = 155,out=-155, min distance=20] node {} (A)
  		(B) edge[loop right, looseness=15, in = 25,out=-25, min distance=20] node {} (B);
\end{tikzpicture}
\caption{The directed graph associated with $J^2$ of the Brusselator. The edges have weights: $1\rightarrow 1: (B-1)^2-A^2B$, $2\rightarrow 2: A^2(A^2-B)$, $1\rightarrow 2: A^2B-B(B-1)$ and $2\rightarrow 1: A^2(B-1)-A^4$.}
\label{J2network}
\end{figure}

\noindent \textit{Graph theoretic formula for the coefficients of a characteristic polynomial.---}Paper \cite{Mincheva2007} gives a graph theoretic formula for the coefficients of characteristic polynomial of the Jacobian matrix of a chemical reaction network, an application of the earlier work of Maybee et al. \cite{Maybee2006} who consider a general square matrix $A$. Following their approach we use the squared Jacobian $J^2$ as an adjacency matrix for a directed graph $\mathcal{G}$. We use a vertex set $V(\mathcal{G}) = \{1,\cdots,n\}$ for an $n$ species interaction network. There is an edge from vertex $i$ to vertex $j$ if $J^2_{ji} \neq 0$. The convention used in \cite{Maybee2006,Mincheva2007} is to only draw self loops if $A_{ii} > 0$, however, for convenience, we will always draw a self loop if $J^2_{ii} \neq 0$. Using these conventions we can draw the directed graph for the Brusselator as shown in Figure \ref{J2network}. We define a cycle $c$ of length $k$ in $\mathcal{G}$ as a series of distinct vertices $\{v_{i_1},\cdots, v_{i_k}\}$ connected by edges $v_{i_1}v_{i_2}, v_{i_2}v_{i_3}, \cdots, v_{i_k}v_{i_1}$.	 For a cycle $c$ we denote $J^2[c] = (J^2)_{v_{i_2}v_{i_1}}(J^2)_{v_{i_3}v_{i_2}}\cdots (J^2)_{v_{i_1}v_{i_k}}$. The Brusselator graph in Figure \ref{J2network} has one cycle of length two with vertices $c_1 := \{v_1,v_2\}$ and two cycles of length one given by the self loops on vertices $v_1$ and $v_2$. A factor $f_k$ of degree $k$ of $\mathcal{G}$ is a collection of pairwise disjoint cycles covering $k$ distinct vertices with $|f_k|$ denoting the number of cycles in $f_k$. The Brusselator has two factors of degree two $f_2 = \{\{c_1\}\}$ and $f_2' = \{\{v_1\},\{v_2\}\}$ and two factors of degree one which are identical to the cycles of length one.

\begin{figure}[t!]
\centering
 \includegraphics[scale=0.45]{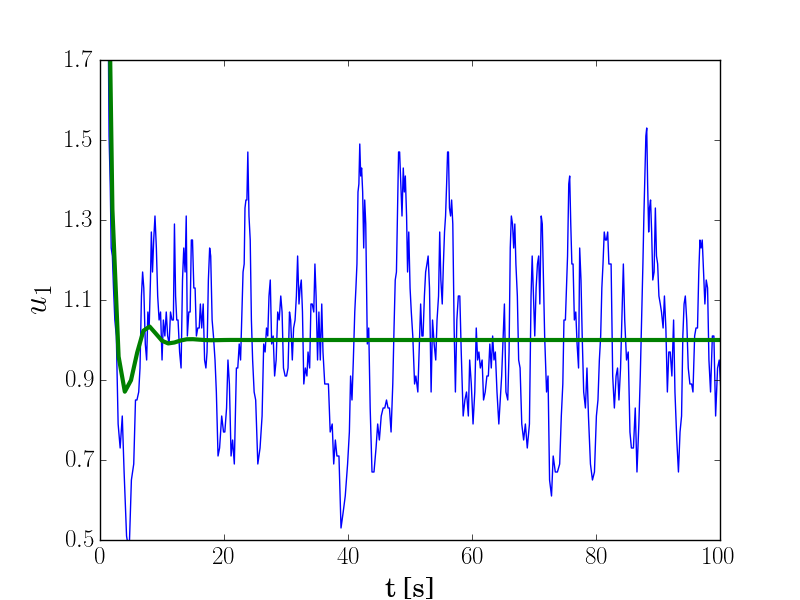}
 \caption{A trajectory of the Brusselator with paramter values $A=1$, $B=1.2$. The smooth green line (colour online) is the solution of the ODE system \eqref{Brusselator} and the oscillating trajectory is the stochastic trajectory.}
 \label{BrussTraj}
\end{figure}

\begin{figure}[t!]
\centering
 \includegraphics[scale=0.45]{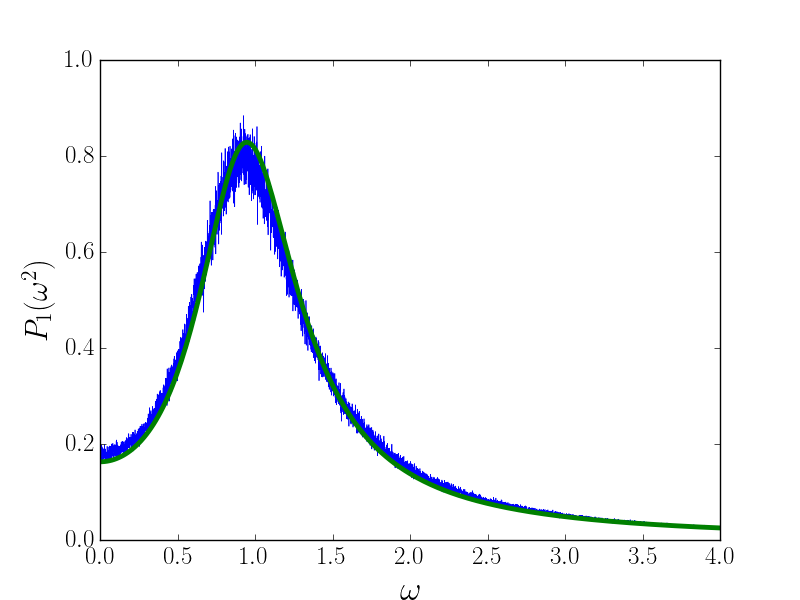}
 \caption{The power spectrum of the stochastic Brusselator for the parameters $A=1$, $B=1.2$. The smooth green line (colour online) represents our theoretical prediction and the oscillating blue line is the average power spectrum of 500 oscillations. We also normalised the theoretical spectrum, equation \eqref{PSpec} and the computational spectrum such that they have unit area.}
 \label{BrussSpec}
\end{figure}

Consider the characteristic polynomial $p(x) = \sum_{i=0}^n a_ix^i$ of a matrix $A$. We can now apply a graph theoretic formula for the coefficients $a_i$, derived in \cite{Maybee2006} and applied to interaction networks in \cite{Mincheva2007},
\begin{equation}
a_{n-k} = \sum_{f_{k} \in \mathcal{G}} (-1)^{|f_{k}|+n+k}\prod_{c \in f_{k}} A[c]\;\;\;\;\;\; k = 1,\cdots,n
\label{ParamEq1}
\end{equation}
where in our example $A = J^2$ and all other quantities are as previously defined.
Therefore, using the cycles and factors we identified in the Brusselator, we can compute the $a_1$ coefficient
\begin{align}
a_1 = (J^2)_{v_1 v_1} + (J^2)_{v_2 v_2} = (B-1)^2 + A^2(A^2 - 2B).
\end{align}
By computing a Sturm chain from the generic polynomial $p(x) = 2x + a_1$ we find that $a_1 < 0$ for the existence of an extremum of $R(\omega^2)$, thus, we re-derived condition \eqref{BrusslCond1}. We simulated the trajectory of the stochastic Brusselator in the parameter regime which satisfies condition \eqref{BrusslCond1} using Gillespie's direct method  \cite{Gillespie1976}, Figure \ref{BrussTraj}, and plotted the power spectrum averaged over 500 repetitions. Our results can be found in Figure \ref{BrussSpec} and show good agreement with our theoretical prediction.

\noindent \textit{Conclusions.---}Resonance in stochastic interaction networks is a well reported phenomenon and a prominent example of how internal stochasticity can lead to oscillatory behaviour. A vital tool to investigate stochastic oscillations is the power spectrum which is traditionally calculated from the Langevin equation. Current methods, however, require detailed knowledge of the underlying stochastic process which can be troublesome to calculate. In this letter we showed how resonance can be understood as a network property, independent of the noise correlations involved. We used Sturm chains to count the number of resonant modes and outlined a graph based method to determine parameter ranges in which stochastic oscillations occur. Future work will seek to extend the application of graph based methods to stochastic spatial systems such as stochastic Turing patterns  in interaction networks.

We would like to thank Eamonn Gaffney for helpful discussions.

\bibliographystyle{unsrt}
\bibliography{PRL-23Feb}

\end{document}